\documentclass{aa}
\usepackage{graphicx,epsfig}
\usepackage{natbib}
\bibpunct{(}{)}{;}{a}{}{,}
\newcommand{\om}{\Omega_{\rm M}}
\newcommand{\ola}{\Omega_\Lambda}
\newcommand{\olx}{\Omega_X}

\def\lsim{\raise0.3ex\hbox{$<$}\kern-0.75em{\lower0.65ex\hbox{$\sim$}}}
\def\gsim{\raise0.3ex\hbox{$>$}\kern-0.75em{\lower0.65ex\hbox{$\sim$}}}                 
\begin{document}

\title{Cosmological parameters from lensed supernovae}
\author{A.~Goobar\inst{1}      \and
        E.~M\"ortsell\inst{1} \and 
        R.~Amanullah\inst{1}\and
        P.~Nugent\inst{2}}
\institute{Department of Physics, Stockholm University, \\
         SCFAB, S--106 91 Stockholm, Sweden \and
Lawrence Berkeley National Laboratory, \\
         1 Cyclotron Road, Berkeley, CA 94720, USA}
\authorrunning{Goobar, M\"ortsell, Amanullah \& Nugent}

\offprints{A. Goobar, ariel@physto.se}

\date{Received ...; accepted ...}

\abstract{
We investigate the possibility of measuring the Hubble constant, the
fractional energy density components and the equation of state
parameter of the ``dark energy'' using lensed multiple images of
high-redshift supernovae.  With future instruments, such as the SNAP
and NGST satellites, it will become possible to observe several
hundred lensed core-collapse supernovae with multiple images. Accurate
measurements of the image separation, flux-ratio, time-delay and
lensing foreground galaxy will provide complementary information to
the cosmological tests based on, e.g., the magnitude-redshift relation
of Type Ia supernovae, especially with regards to the Hubble parameter
that could be measured with a statistical uncertainty at the one
percent level. Assuming a flat universe, the statistical uncertainty
on the mass density is found to be $\sigma^{\rm stat}_{\om} \lsim
0.05$.  However, systematic effects from the uncertainty of the lens
modeling are likely to dominate. E.g., if the lensing galaxies are
extremely compact but are (erroneously) modeled as singular isothermal
spheres, the mass density is biased by $\sigma^{\rm syst}_{\om} \sim
0.1$.

We argue that wide-field near-IR instruments such as the one proposed
for the SNAP mission are critical for collecting large statistics of
lensed supernovae.}
\maketitle
\section{Introduction}
Gravitational lensing of high-$z$ objects has been used in the past
as a tool for deriving cosmological parameters with mixed
success. While the fraction of quasars with multiple images and the
distribution of image separations may be used to probe the vacuum
energy density, $\Omega_\Lambda$ \citep{turner90,sef}, the method
suffers from severe systematic uncertainties related to the lens
modeling and the results are as of yet inconclusive
\citep{fukugita92,maoz,kocha96,park,chiba}.  On the other hand, the
method proposed by \citet{refsdalb,refsdalc} using time delay
measurements of multiple imaged quasars to constrain $H_0$ has
provided results that are in good agreement with independent
techniques,
\citep[see e.g.][]{koopmans99,brown}. Thus, distances derived from
geometrical measurements of lensed high-redshift sources may be a
viable way to gain further knowledge of cosmological parameters
although there are still unresolved issues concerning the modeling of
the lenses. E.g. in a recent paper \citep{kocha02}, five gravitational
lens systems are analyzed and it is suggested that, unless the dark
matter distributions in the lensing galaxies are rather compact, the
derived value of $H_0$ is too low 
in comparison with the local measurements.

In this note we investigate how to use multiply imaged high-redshift
supernovae (SNe) to constrain the Hubble parameter, the mass and dark
energy density of the universe, $\om$ and $\olx$, as well as the
equation of state parameter, $w_0 = {p_X \over \rho_X}$, which we
assume is constant for $z\lsim 5$. SNe are well suited for this
technique because of the expected high rate at high-redshifts and most
importantly because of their well known lightcurves. In particular,
the (rest-frame) optical lightcurves of core-collapse (CC) SNe
show fast rise-times, typically about 1 week long. Thus, the time
difference between images can be measured to better than one day's
precision.

Another possibility is the use of the UV shock-breakout. As has been seen
in SN 1987A and modeled in \citet{ensman} the shock breakout can
serve as a time stamp with a precision of just minutes. The entire UV
flash occurs over a period of minutes to several hours in the rest-frame of
the SN depending on the nature of the progenitor.  

Wide field optical and NIR deep surveys, such as the planned SNAP 
satellite \citep{snap}, have the potential to discover $\sim10^6$ CC SNe. While
these SNe are sometimes regarded as a ``background'' for the primary
cosmology program based on Type Ia SNe, we argue that lensed SNe of
{\em any} kind may also provide useful information on cosmological
parameters.

In \citet{holz} the number of multiply imaged CC SNe up to 
$z < 2$ were estimated by simply scaling the Type Ia rate by 
a factor of 5. In this work we extend the considered redshift
up to $z=5$ using a SN rate calculation derived from the star
formation history. Further, we take into account the NIR
wide field instrumentation in the current design of the SNAP 
satellite. 

Using simple toy-models, we investigate the accuracy of the
strong lensing technique to improve our knowledge of 
cosmological parameters. Our observables are the source redshift, the redshift of
the lensing galaxy, the image-separation, $\Delta\theta$, the
time-delay, $\Delta t$, and the flux-ratio, $r$.  We derive the
relation of our observables and cosmological parameters for different
matter distributions in order to investigate the sensitivity to the
choice of lens model.  We speculate, that if the measurement is done
using a very large sample of lensed systems, useful bounds on the
cosmological parameters can be found in spite of large uncertainties
in the lens model.

We have used the SNOC Monte-Carlo simulation package \citep{snoc}
to estimate the rate and measurable quantities of multiple image SNe, e.g. the 
distribution of time-delays, image-separations and flux-ratios. We also simulate 
extinction by dust both in the host galaxy of the SN and in the foreground lensing
galaxy.

While about
half of the known lensed systems exhibit more images \citep[see e.g.][]{keeton98}, 
this work is limited
to spherically symmetric lensing systems producing only two or ring-like images. 
Systems with more lensed images
are potentially very interesting as they provide more measurables which
can be used to constrain the lens model. 

\section{Modeling of galaxy halos}
In this work, we confine our study to a very simple class of spherically symmetric
lens models with projected, two-dimensional lens potential \citep{sef} 
\begin{equation}
        \label{eq:psi}
        \psi (r) =k r^\alpha , 
\end{equation}
where $\alpha\le 1$ to assure that we get multiple images and that the
surface density falls at large radii.  The two limiting cases of this
class of models are the {\em Singular Isothermal Sphere} (SIS) where
$\alpha =1$ and the point-mass for which $\alpha =0$ and $k\alpha =1$.
A drawback with this choice of single-slope models is that we need
steep halo models in order to produce multiple images. A slightly more
general treatment would include double slope models like the
isothermal sphere with a core (ISC) or the NFW density profile
\citep{nfw}.  Also, since many multiple image systems have more than
two images, the ellipticity of the lensing galaxy should be included
in the model.  That is, since we assume a very limited class of lens
models, we underestimate the systematic effects from the lens
modeling. However, we also use a very limited set of observables. In
four-image systems, we will have ten observables to use in the fit
(four image positions, three flux-ratios and three time-delays) in
comparison to the three observables used in this study. Still, there
is no doubt that we underestimate the errors in the lensing potential
while restricting our study to lens potentials of the form given by
Eq.~(\ref{eq:psi}). As a first step, we compare results for the two
extreme cases of point-masses and SIS density profiles.

\subsection{Point-masses}

A light ray which passes by a point-mass $M$ at a minimum distance
$\xi$, is deflected by the ``Einstein angle''
\begin{equation}
  \hat\alpha=\frac{4M}{\xi}=\frac{2R_S}{\xi},
\end{equation}
where $R_S=2M$ is the Schwarzschild radius.
Using a characteristic length $\xi_0$ in the lens plane given by
\begin{equation}
  \xi_0=\sqrt{2R_S\frac{D_dD_{ds}}{D_s}}=\beta_0 D_d,
\end{equation}
\begin{equation}
  \beta_0=\sqrt{2R_S\frac{D_{ds}}{D_dD_s}},
\end{equation}
where $D_{ds}, D_d$ and $D_s$ are angular diameter distances between
deflector and source, deflector and observer, and source and
observer respectively. The lens equation in dimensionless form is
\begin{equation}
  y=x-1/x,
\end{equation}
which has two solutions,
\begin{equation}
  \label{eq:solution}
  x_{1,2}=\frac{1}{2}\left (y\pm \sqrt{y^2+4}\right ),
\end{equation}
i.e., one image on each side of the lens. The image separation is given by
\begin{equation}
  \Delta\theta=\beta_0\sqrt{y^2+4}.
\end{equation}
The magnification is given by
\begin{equation}
  \mu =\left (1-\frac{1}{x^4}\right )^{-1},
\end{equation}
which can be combined with Eq.~\ref{eq:solution} to give
\begin{equation}
  \mu_{1,2}=\pm\frac{1}{4}\left [\frac{y}{\sqrt{y^2+4}}+\frac{\sqrt{y^2+4}}{y}
            \pm 2\right ].
\end{equation}
The ratio $r$ of the absolute values of the magnifications (flux-ratio) 
for the two images produced by a point-mass lens is given by,
\begin{equation}
  r=\left |\frac{\mu_1}{\mu_2}\right |=\left [\frac{\sqrt{y^2+4}+y}
    {\sqrt{y^2+4}-y}\right ]^2,
\end{equation}
and
\begin{equation}
  y=r^{1/4}-r^{-1/4}.
\end{equation}
Thus, we can write the image separation in terms of the flux-ratio as
\begin{equation}
  \Delta\theta=\beta_0(r^{1/4}+r^{-1/4}).
\end{equation}

The time delay between images is in the general case given by
\begin{equation}
  \Delta t(\vec y)=\xi_0^2\frac{D_s}{D_dD_{ds}}(1+z_d)
                    \left [\phi(\vec x^{(1)},\vec y)-
                    \phi(\vec x^{(2)},\vec y)\right],
\end{equation}
where $\phi(\vec x,\vec y)$ denotes the so called \emph{Fermat potential}.
In the case of a point-mass lens we have
\begin{equation}
  \Delta t=2R_S(1+z_d)\tau(y),
\end{equation}
where
\begin{equation}
  \tau(y)=\frac{1}{2}y\sqrt{y^2+4}+\ln\frac{\sqrt{y^2+4}+y}{\sqrt{y^2+4}-y}.
\end{equation}
We can now express $R_S$ and $y$ in our observables to get
\begin{equation}
  \Delta t=\frac{\Delta\theta^2}{2}\frac{D_dD_s}{D_{ds}}(1+z_d)
  \frac{\sqrt{r}-1/\sqrt{r}+\ln{r}}{\sqrt{r}+1/\sqrt{r}+2}.
 \label{eq:poi}
\end{equation}

\subsection{Singular isothermal sphere (SIS)}
The density profile for the SIS halo, which is frequently used in 
gravitational lensing analysis, is given by
\begin{equation}
  \rho_{{\rm SIS}}(r)=\frac{v^2}{2\pi}\frac{1}{r^2}.
\end{equation}
Here, $v$, the only free parameter of the model, is the line-of-sight
velocity dispersion.  The simplicity of the SIS density profile allows
analytical solutions for the many quantities related to gravitational
lensing \citep{turner84}.  The deflection angle is computed to be
\begin{equation}
  \hat\alpha (\xi)=4\pi\left(\frac{v}{c}\right) ^2\frac{|\xi |}{\xi} 
  =:\alpha_0\frac{|\xi |}{\xi},
\end{equation}
where $\alpha_0$ is defined as the magnitude of the deflection.
That is, the magnitude of the deflection angle is independent of the impact 
parameter. 
Choosing
\begin{equation}
  \label{eq:xi0}
  \xi_0=4\pi\left (\frac{v}{c}\right )^2\frac{D_dD_{ds}}{D_s},
\end{equation}
the lens equation can be written
\begin{equation}
  \label{eq:sislens}
  y=x-\frac{x}{|x|}.
\end{equation}
Considering $y>0$, there are two images for any $y<1$; at $x=y+1$ and
$x=y-1$, i.e., on opposite sides of the lens center.
The image separation is given by 
\begin{equation}
  \label{eq:delthe}
  \Delta\theta =2\alpha_0\frac{D_{ds}}{D_s},
\end{equation}
and the magnification for an image at $x$ is
\begin{equation}
  \label{eq:sismag}
  \mu =\frac{|x|}{|x|-1}.
\end{equation}
From Eq.~(\ref{eq:sismag}) we see that the ratio of the absolute values of
magnifications of the images is
\begin{equation}
  r=\frac{1+y}{1-y}.
\end{equation}

The time delay for the two images is
\begin{equation}
  \Delta t=\left [4\pi\left (\frac{v}{c}\right )^2\right ]^2
            \frac{D_dD_{ds}}{D_s}(1+z_d)2y,
\end{equation}
which can be expressed in terms of our observables as
\begin{equation}
  \Delta t=\frac{\Delta\theta^2}{2}\frac{D_dD_s}{D_{ds}}(1+z_d)
  \frac{r-1}{r+1}.
 \label{eq:sis}
\end{equation}
We see that only the $r$-dependent part differs between the point and the 
SIS case. It is reassuring to see that they
only differ by at most 6\% up to a flux ratio of four. Thus, the derived value
of the cosmological parameters are only weakly sensitive to the 
choice of point or SIS model.

\section{Monte-Carlo simulations}

Since we in principle only are able to follow infinitesimal light-beams
in SNOC, we have to use some approximations when trying to get
information on multiple image systems. The main approximation is that
we assume that in cases of strong lensing, the effects from one close
encounter is dominant, i.e., the one-lens approximation. With this
simplification, we can use the information from the magnification to
derive quantities for systems with finite separations. In order to do
this, we need to be able to derive analytical relations between the
magnification and the image-separation and so forth. Here we show how
this is done for the case of SIS lenses.  

First, we concentrate on primary images.  Studying
Eqs.~(\ref{eq:sislens}) and (\ref{eq:sismag}), we see that multiple
imaging occurs whenever $\mu_1>2$ and that the magnification of the
second image is given by
\begin{equation}
  |\mu_2|=\mu_1-2.
\end{equation}
Using Eq.~(\ref{eq:delthe}) and the fact that $y=1/(\mu_1-1)$, we can write the 
time-delay for a SIS lens as 
\begin{equation}
  \Delta t=\left [4\pi\left (\frac{v}{c}\right )^2\right ]^2
            \frac{D_dD_{ds}}{D_s}(1+z_d)\frac{2}{\mu_1-1}.
\end{equation}
Therefore, in order to compute the quantities of interest, we need to
pick a lens redshift and velocity dispersion from some reasonable
distributions for every case where $\mu_1>2$. The distributions
will in general be functions of the cosmology, the mass distribution
of the lenses and the source redshift.
The differential probability for multiple imaging is in the general case
given by
\begin{equation}
  dP\propto\sigma (z_d,z_s)\frac{dn}{dM}dM (1+z_d)^3 dV,
\end{equation}
where $\sigma (z_d,z_s)$ is the cross-section for multiple imaging and $n$ is
the comoving number density of lenses. For SIS lenses, the cross-section
is given by
\begin{equation}
  \label{eq:cc}
  \sigma (z_d,z_s)\propto\left(\frac{v}{c}\right)^4D_{ds}^2.
\end{equation}
Since $dV\propto D_d^2\frac{dt}{dz_d}dz_d$ and $n$ is independent of the lens
redshift we can use
\begin{equation}
  dP(z_d)\propto D_{ds}^2D_d^2(1+z_d)^3\frac{dt}{dz_d},
\end{equation}
as our probability distribution for $z_d$. The probability distribution for $v$
is given by
\begin{equation}
  dP(v)\propto \frac{dn}{dM}\frac{dM}{dv}\left(\frac{v}{c}\right)^4dv.
\end{equation}
Following \citet{bergstrom}, we derive a galaxy mass
distribution, $dn/dM$, by combining the Schechter luminosity function
\citep[Eq.~5.129]{book:Peebles},
\begin{eqnarray}
  dn&=&\phi(y)dy , \nonumber \\
  \phi(y)&=&\phi_* y^\alpha e^{-y} , \label{eq:Schechter} \\
  y&=&\frac{L}{L_*} \nonumber ,
\end{eqnarray}
with the mass-to-luminosity ratio \citep[Eq.~3.39]{book:Peebles}, 
\begin{equation}
  \label{eq:masstolum}
  \frac{M}{M_*}=y^{1/(1-\beta)}.
\end{equation}
Using Eq.~(\ref{eq:Schechter}), we find that
\begin{eqnarray}
  \frac{dn}{dM}&\propto&y^\delta e^{-y} , \\
  \delta&=&\alpha-\frac{\beta}{1-\beta} .
\end{eqnarray}
Combining the Faber-Jackson relation
\begin{equation}
  \frac{v}{v_{*}}=y^{\gamma} ,
\end{equation}
where $y$ is defined in Eq.~(\ref{eq:Schechter}), with the
mass-to-luminosity ratio, Eq.~(\ref{eq:masstolum}), we can relate the
velocity dispersion to mass by
\begin{equation}
  \frac{v}{v_*}=\frac{M}{M_*}^{\gamma (1-\beta)}.
\end{equation}
In this paper we use $v_*=220$ km s$^{-1}$, $\alpha =-1.07$, $\beta
=0.2$ and $\gamma =0.25$. The mass normalization is calculated
assuming that the entire mass of the resides in galaxy halos
\citep{bergstrom}. 

A higher value of $v_*$ would result in a larger number of wide
separation lenses since the cross-section for multiple imaging scales
as $\sigma\propto v^4$ [see Eq.~(\ref{eq:cc})] and the image
separation as $\Delta\theta\propto v^2$ [Eq.~(\ref{eq:delthe})].
Varying the galaxy mass distribution will also have an effect on the
characteristics of the lensed events. However, since the mass and
velocity dispersion are not variables in the fitting of cosmological
parameters (see Sect.~\ref{sec:fitting}), any changes in these
distributions will only have a marginal effect on the results obtained
in this paper.
        
\section{Characteristics of lensed SN events}

We have simulated $1.1 \cdot 10^6$ CC SNe in the redshift range $0 \le
z \le 5$ using a SIS model for the galaxy density profiles. The
redshift dependence of the SN rates, the relative fraction among the
various types of CC SNe (Ib/c, IIL, IIn, IIP, 87a-like), peak
magnitudes and intrinsic dispersion were simulated following the
prescriptions in \citet{dah99} and
\citet{dah02}. Using SN rate predictions derived 
from the star formation history \citep{chugai,sullivan}, our simulated
sample corresponds approximately to the predicted number of CC SN
explosions ($z \le 5$) in a period of 3-years in a 20 square degree
field, i.e., $\sim 5.1$ SNe year$^{-1}$arcmin$^{-2}$.  In the
simulations we have assumed a flat cosmology with $\om$ = 0.3 and
$\Omega_{\Lambda}$ = 0.7, and a Hubble constant $H_0 = 65$ km
s$^{-1}$Mpc$^{-1}$. We view these rates on the conservatively low
side given the recent work by
\citet{lanzetta} who finds that the SFR plausibly increases monotonically
with redshift through the highest redshifts observed. 

\begin{figure}[t]
  \centerline{\hbox{\epsfig{figure=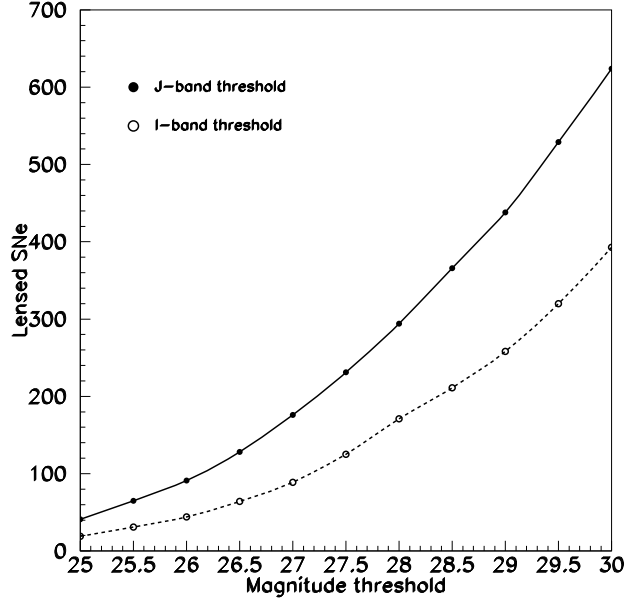,width=0.5\textwidth}}}
  \caption{Number of multiple image SNe in the simulated sample vs magnitude threshold 
   for the faintest image in I and J-band.}
  \label{fig:threshold}
\end{figure}

Out of the $1.1 \cdot 10^6$ simulated CC SNe, 2613 were multiply
lensed. Fig.~\ref{fig:threshold} shows the number of detectable SNe as
a function of the peak-brightness threshold in I and J-bands. The
threshold refers to the {\em faintest} of the two images.

In the following analysis, we consider two possible data samples 
(all magnitudes are in the Vega system):
 
\begin{itemize}
\item case A: only SNe with an I-band peak magnitude of the faintest image 
of m$_{\rm I} \le 28.5$ 

\item case B: SNe where {\em either} the I-band peak brightness {\em or} the 
J-band brightness falls below 28.5 mag (i.e. m$_{\rm I}$ or m$_{\rm J}
\le$ 28.5)
\end{itemize}

Thus, only 211 (case A) and 366
(case B) lensed CC SNe were considered for cosmology fits. Additional
potential SNe discovered in other bands are thus not considered in
this study. For an instrument like SNAP (wavelength sensitivity
0.4-1.7 $\mu$m), revisiting each field in at least one optical or NIR
band every 6 (or less) hours, we expect a very high detection
efficiency. E.g., assuming that the SNe were discovered 0.5 magnitudes
below peak brightness, SNAP would be able to make a 5$\sigma$ detection
every four days in each of three (or more) filters: I, Z , J and
possibly others.

In the simulations we have considered the possibility that one or both
images suffer extinction. We assumed differential extinction
properties as in \citet{Cardelli}, with $R_V=3.1$ and a mean-free path
of 1 kpc for V-band photons in the host and lensing galaxy. Details of
the simulation procedure can be found in \citet{dust,snoc}.

\begin{figure}[t]
  \centerline{\hbox{\epsfig{figure=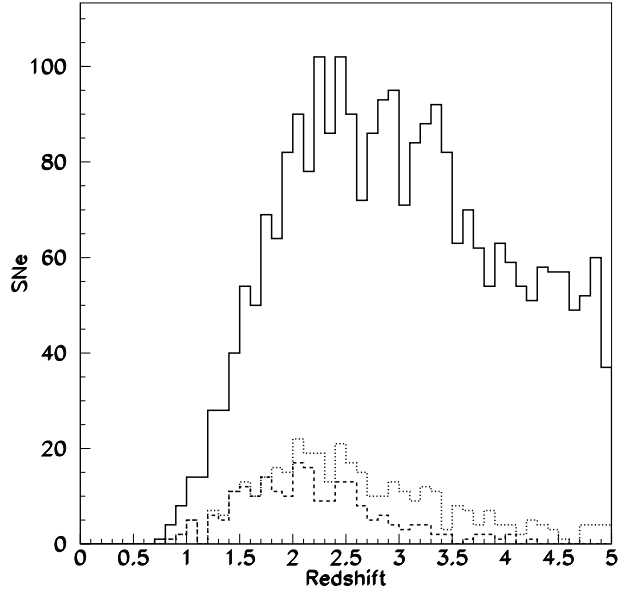,width=0.5\textwidth}}}
  \caption{Redshift distribution of lensed SNe up to $z=5$, solid
  line. The dotted line shows SNe for which the SN system satisfies
  the case B criteria. The dashed lines shows case A. A SIS lens
  profile was used. The lensing galaxy is typically at a redshift $\sim z_s/2$}  \label{fig:multi_zs}
\end{figure}

\begin{figure}[t]
  \centerline{\hbox{\epsfig{figure=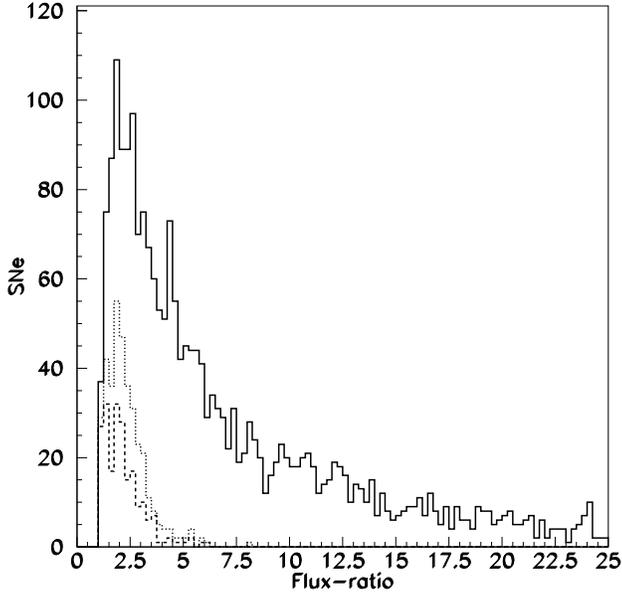,width=0.5\textwidth}}}
  \caption{Flux-ratio of SN images. The dotted line shows SNe for
  which the SN system satisfies the case B criteria. The dashed line
  shows case A. A SIS lens profile was used.} \label{fig:multi_ratio}
\end{figure}

\begin{figure}[t]
  \centerline{\hbox{\epsfig{figure=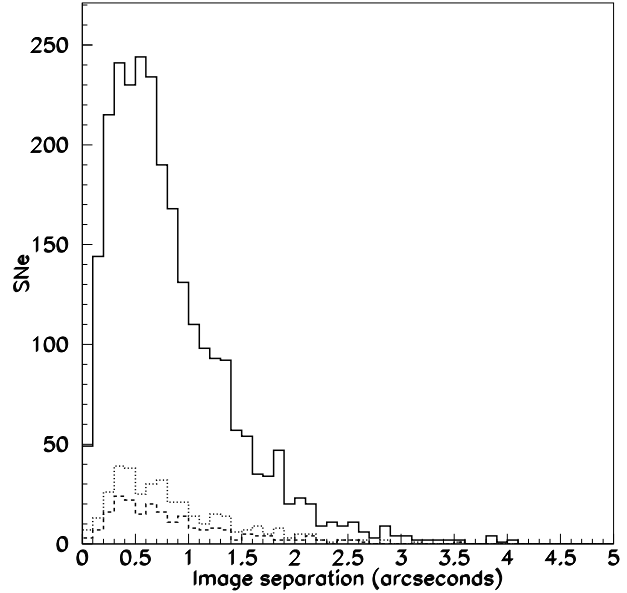,width=0.5\textwidth}}}
  \caption{Image separation in arcseconds. The dotted line shows SNe
  for which the SN system satisfies the case B criteria. The dashed
  line shows case A. A SIS lens profile was used.} 
  \label{fig:multi_ismep}
\end{figure}

\begin{figure}[t]
  \centerline{\hbox{\epsfig{figure=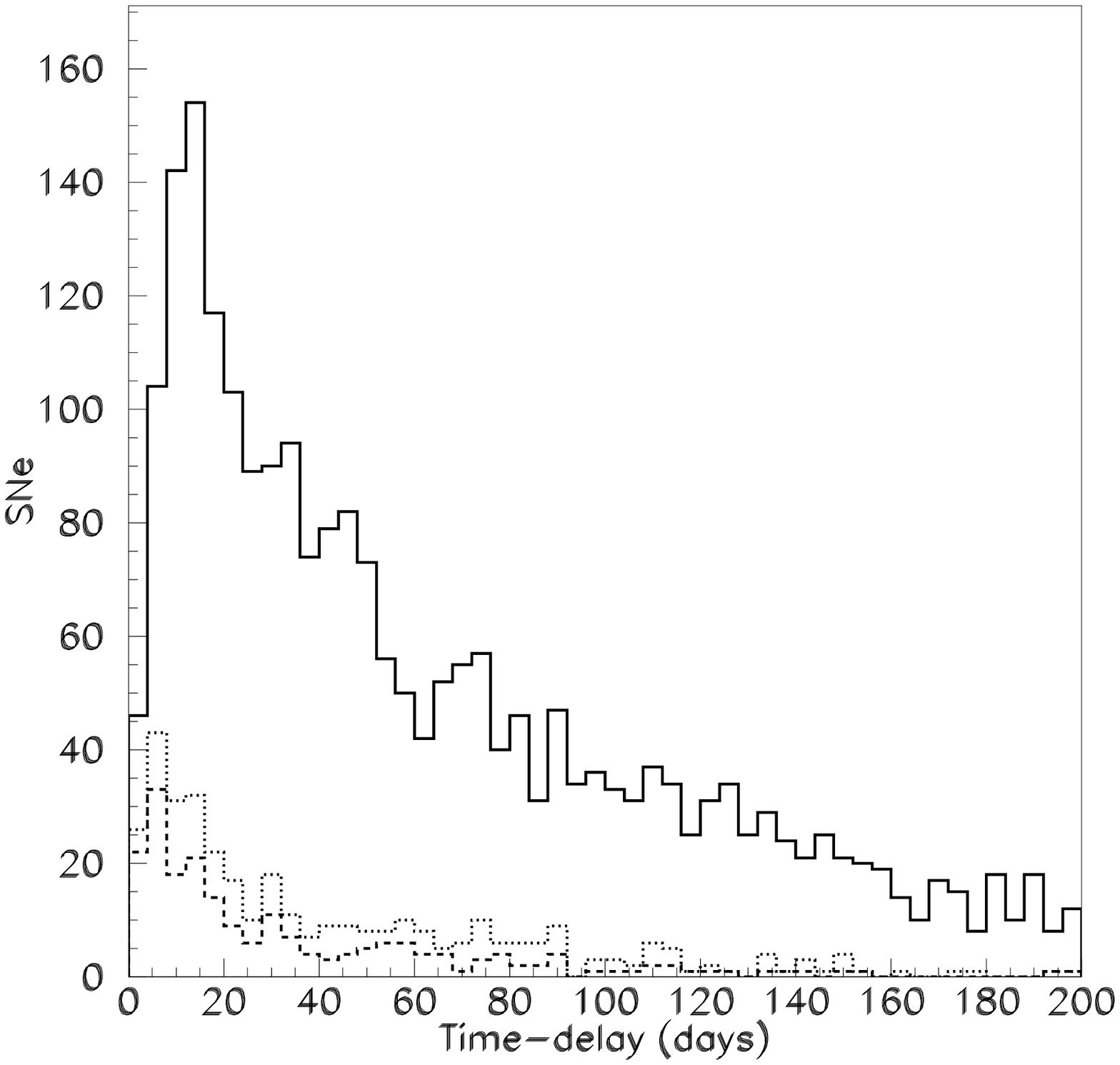,width=0.5\textwidth}}}
  \caption{Time delay between the SN images in days. The dotted line
  shows SNe for which the SN system satisfies the case B
  criteria. The dashed line shows case A. A SIS lens profile was used.} 
  \label{fig:multi_tdelay}
\end{figure}

In Figs. \ref{fig:multi_zs}--\ref{fig:multi_tdelay}, we show the
distribution of observables for the multiple-image SNe; source
redshifts, image flux-ratios, image separations and
time-delay between the images. The dashed lines show the subset 
of the data fulfilling case A SNe
while the dotted curves correspond to case B.

The salient features of the events are: images of the same source with average 
redshift $z_s \approx 2$ and the lensing galaxy typically at $z_d \sim z_s/2$.
The SN images are  separated by
$\sim$\,0.5 arcseconds and a few weeks apart. With the imposed minimal brightness
criteria, the images differ typically by less than one astronomical magnitude.
Clearly, a very clean signature to search for, especially with space observations
free of atmospheric blurring.

\section{Fitting cosmological parameters}\label{sec:fitting}

Next, we estimate the target statistical uncertainty of a 3 year
mission studying a 20 square degree patch of the sky down to 28.5 
peak I-band (and J-band) magnitudes. We perform a maximal likelihood
analysis of the multiply lensed systems with $h, \om$ and $\ola$ as
free parameters where $h$ is the dimensionless Hubble parameter
defined by $H_0 = h
\cdot$100 km s$^{-1}$Mpc$^{-1}$.  With $\vec{\phi}=(h,\om,\ola)$ and
$f(r)$ being the $r$-dependent factors in Eqs.~(\ref{eq:poi}) and
(\ref{eq:sis}) for point mass lenses and SIS respectively we define
the ``experimental'' ($R_{\rm exp}$) and cosmology dependent (${\cal
R}$) quantities:

\begin{eqnarray*}
 R_{\rm exp} = 2 {\Delta t \over {\Delta \theta}^2}\left({1 + z_d \over f(r)}\right)^{-1}  {\rm and} \\
{\cal R}(\vec{\phi})  = {D_{d} \cdot D_{s}  \over D_{ds}}. 
\end{eqnarray*}

The maximum likelihood analysis performed over the simulated SNe with index
$i=$1..N assumes Gaussian probability density functions for
the observables, 
\begin{equation}
{\cal L} = \prod_i {1 \over \sqrt{2 \pi \sigma}} e^{-(R_{\rm exp}^i -
 {\cal R(\vec{\phi}}))^2/2\sigma^2}, 
\label{eq:maxl}
\end{equation}
where the standard deviation $\sigma$ contains the propagated uncertainties of the 
(independent) observables $\vec{x} = (\Delta t, \Delta \theta, r, z_d)$:
\begin{equation}
 \sigma^2 = \sum_{j=1}^4{\left({\partial R_{\rm exp} \over \partial x_j}  \sigma_{x_j} \right)^2}. 
 \label{eq:erexp}
\end{equation}
For simplicity, we have assumed that all SNe are measured with the same
precision.

While we will consider more pessimistic scenarios later, in this section we assume that 
the uncertainties $\sigma_{x_j}$ of the four observables are:
\begin{eqnarray}
\left\{\begin{array}{lcl}
\sigma_{\Delta t} & = & 0.05 {\rm \ days} \\ 
\sigma_{\Delta \theta}   & = & 0.01'' \\
\sigma_{r}        & = & 0.1    \\
\sigma_{z_d}      & = & 0.001 
\end{array}
\right. 
 \label{eq:errors}
\end{eqnarray}

The outstanding precision in the time delay between the SN images
is consistent with the estimates of the accuracy expected for SNIa
lightcurves studied with simulations by the SNAP collaboration.

The image separation uncertainty corresponds  to 0.1 pixel of
the proposed SNAP instrument. This estimate  could even improve
considering that all of the SN images 
along the lightcurve can be co-added to get $\gsim 40 \sigma$ signals 
on the individually lensed SNe. E.g. in \citet{wfpc} it is stated that
with HST/WFPC one could reach 0.02 pixel precision on reasonable bright stars.

In estimating $\sigma_{r}$ we have assumed that the differential
extinction of the lensed SNe will not largely exceed what has been
measured for a set of 23 gravitational lens galaxies in \citet{falco}.
The median extinction for those systems ($z_d\lsim 1$) was found to be
$E(B-V)\sim 0.05$ mag.

\subsection{The $\om-\ola$ plane}
\begin{figure}[t]
  \centerline{\hbox{\epsfig{figure=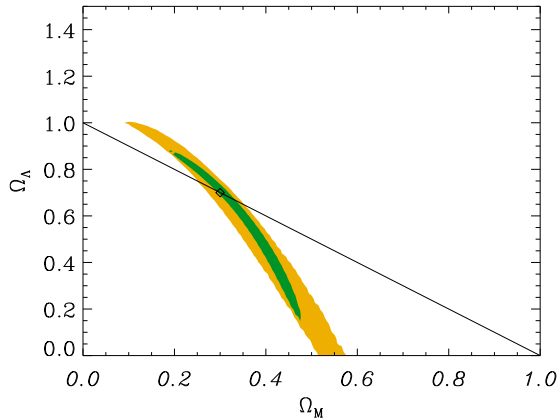,width=0.5\textwidth}}}
  \caption{68 \% CL region in the $\om-\ola$ projection of a three
  parameter fit $(\om,\ola,h)$.
  The 366 two-image events fulfilling criteria B. The
  dark (green) region shows the 68 \% CL region that would result if $h$
  would be exactly known from independent measurements. The line shows
  $\om+\ola=1$, i.e. a flat universe. The diamond shows the value that
  was used in the simulation.}  
\label{fig:omox_i29}
\end{figure}

While ${\cal R}\propto h^{-1}$, it varies only weakly with $\om$ and
$\olx$. This can be appreciated from the projected 68 \% confidence
level (CL) region from a 3-parameter fit ($\om,\ola,h$) to the
simulated data-set B in Fig.~\ref{fig:omox_i29}. From now on we
concentrate on the case B sample. The areas covered by the contour
regions of case A scale approximately as $\sqrt{N_B/N_A}$. It should
be noted that the shape and inclination of CL region is very different
from what is found from the magnitude-redshift test \citep{goo95}. We
also show in dark the constraints in the $\ola-\om$ plane if $h$ is
known (exactly). While little information can be gained from $\om$ or
$\ola$ {\em independently}, the figures indicate that there is only a
very limited range in $\om$ and $\ola$ for which the CL region is
consistent with a flat universe, $\om + \ola = 1$. Thus, from now
on, we concentrate on the cases where the geometry is assumed to be
known from, e.g., CMB anisotropy measurements.

\subsection{The $\om-h$ plane}

\begin{figure}[t]
  \centerline{\hbox{\epsfig{figure=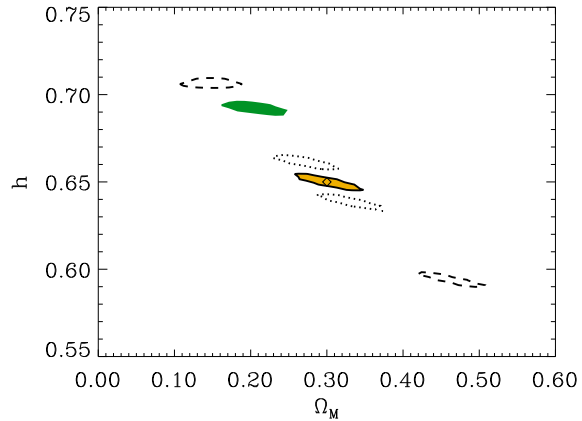,width=0.5\textwidth}}}
   \caption{The light (yellow) shaded region bounded by the solid line shows 68 \% CL region of $\om-h$ fit from 366
   two-image events where the peak brightness of the faintest image
   fulfills criteria B. 
   A flat universe was assumed. The dark (green) shaded region shows the bias introduced
   if the fit is done (erroneously) assuming a point-mass
   lens model. The dotted and dashed curves show the effect of a systematic error in the lens model, according
   to Eq.~\ref{eq:esyst}}  \label{fig:omh0_flat}
\end{figure}

With large statistics, such as the case in our simulations, it is
feasible to constrain the possible values of $\om$ if a flat universe
is assumed. Fig.~\ref{fig:omh0_flat} shows the 68 \% CL-region of the
$\om-h$ plane for case B. The dark shaded region shows the systematic
effect introduced by fitting the SIS simulated data with the
assumption of point mass lenses.  The fit in the light shaded region (solid
curve) of
Fig.~\ref{fig:omh0_flat} yields $h = 0.65^{+0.006}_{-0.003}$; $\om =
0.30^{+0.04}_{-0.05}$. The uncertainties are 68 \% CL for a two
parameter fit, i.e., 1.51 $\sigma$.  Introducing the bias due to the
wrong lens model, we found the fitted central values to be $h = 0.69$
and $\om = 0.19$.
To further test potential bias in the cosmological parameters 
from systematic uncertainties in the lens model we have tested
adding an ad-hoc offset with a different dependence on the
image ratio:  
\begin{equation}
 R_{\rm exp}^{b} = R_{\rm exp} (1 +\epsilon r^2) ,
 \label{eq:esyst} 
\end{equation}

where $\epsilon=\pm 0.001$ and $\pm 0.005$  and the considered uncertainties in the 
fit are as in Eq.~\ref{eq:errors}. The effect is 
shown as  dotted and dashed curves 
in Fig.~\ref{fig:omh0_flat}. Note, however, that in these examples
we are assuming that {\em all} lensing galaxies are wrongly modeled. 
Thus, in that sense, it should be a conservative estimate of the systematic
uncertainties.

It is interesting to note that if the Hubble parameter and
the energy densities are known to good accuracy from other cosmological tests, it should be
possible to put very useful constraints on the average lens properties.

\subsection{The $\om-w_0$ and $h-w_0$  planes}
\begin{figure}[t]
  \centerline{\hbox{\epsfig{figure=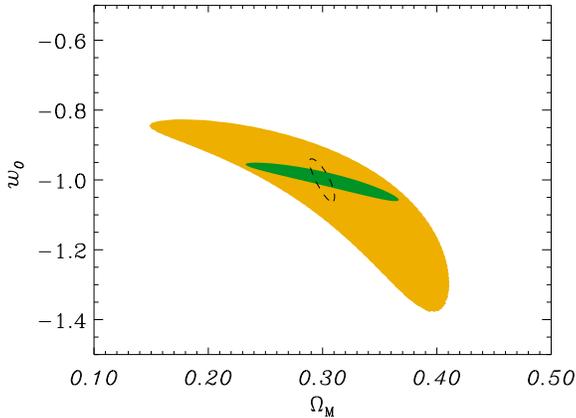,width=0.5\textwidth}}}
   \caption{68 \% CL region of $\om-w_0$ fit from the 366 events fulfilling 
    criteria B. A flat universe was assumed.  The
  dark (green) region shows the smaller confidence region that would result if $h$
  would be exactly known from independent measurements. The dashed line shows the
 expected statistical uncertainty from a 3 year SNAP data sample of Type Ia SNe. }  
\label{fig:omw0}
\end{figure}

Next, we investigate the potential of this method in setting limits in
the $\om-w_0$ parameter plane.  Fig.~\ref{fig:omw0}
 shows that upper
limits on the equation of state parameter of dark energy\footnote{
Shown in the figure is the 2D projection of
the 3-parameter fit ($h$, $\om$, $w_0$).}, $w_0 < -0.8$ may be
derived from the considered data sample. Meaningful limits on the
possibility of $w_0<-1$ could be derived, especially if an independent estimate of
$H_0$ is used as a prior in the fit.

\begin{figure}[t]
  \centerline{\hbox{\epsfig{figure=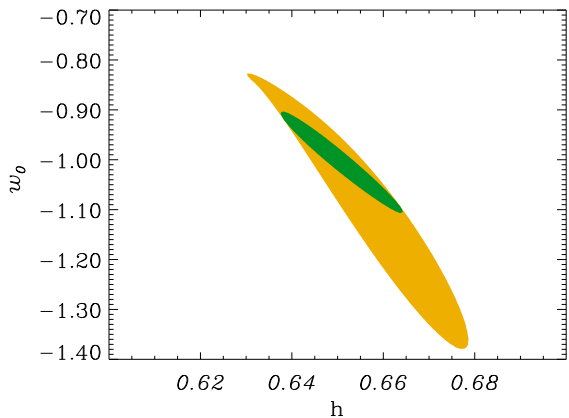,width=0.5\textwidth}}}
   \caption{68 \% CL region of $h-w_0$ fit from 366 
   two-image events in case B. A flat universe was assumed.
 The dark (green) region shows smaller confidence region that would result if $\om$
  would be exactly known from independent measurements.
}  
\label{fig:w0h}
\end{figure}

Due to the different reshift distributions considered, the shape of
the CL-regions differ from what is expected from the Hubble diagram of
Type Ia SNe for the SNAP satellite \citep{goliath}. In particular, if
the Hubble parameter is further constrained by some independent
method, the estimates of $w_0$ from the strong lensing data becomes
comparable in precision the limits that may be derived from Type Ia
SNe, the dashed countour line in Fig.~\ref{fig:omw0}.  The other 2D
projection, the one onto the $w_0-h$ plane, is shown in
Fig.~\ref{fig:w0h}. If the wrong halo model is used in the 
fitting procedure, i.e., a point-mass halo model instead of SIS, 
a 5 \% bias is introduced in the estimate of $w_0$.

\section{Summary and conclusions}
Strongly gravitationally lensed SNe could be detected in large numbers in planned wide
field, deep, SN search programs, probably on the order of several hundred. In particular, we argue
in favor of large NIR imagers in space missions like SNAP. 

Lensed SNe are potentially interesting as they provide independent
measurements of cosmological parameters, mainly $H_0$, but also the
energy density fractions and the equation of state of dark energy. The
results are independent of, and would therefore complement the Type Ia
program. At the faint limits considered in this note, several quasars
per square arcminute are expected \citep{quasars}. Thus, we expect
that an instrument like SNAP would find several hundred multiply
imaged QSOs, in addition to the strongly lensed SNe. Thus, the
statistical uncertainty could become smaller than what we have
considered here.

While the systematic uncertainties remain a source of concern, we show
that the simplest spherically symmetric models introduce moderate
biases ($\sigma^{\rm syst}_{\om} \lsim 0.1$) , at least as long as multiple images with similar fluxes are
considered.

While projections of different SNe could in principle be
interpreted as a lensed SN the two scenarios may be distinguished.
The signatures of CC SNe are unique. Crudely, the lightcurves
are a product of the progenitor mass, the mass loss during its
evolution off the main sequence, the amount of radioactive Ni
synthesized during the explosion and the kinetic energy imparted to
the ejecta. In addition, the environments the SNe explode in
(the density and structure of their local interstellar medium), often
play a significant role in what we eventually see of the CC
event. These differences have given rise to all the different
classifications of these events we currently have; Type IIP, IIL, Ib,
Ic, IIn, etc. Given all this diversity it makes it quite easy to
distinguish one event from another and to not confuse a lensing event
with a coincident CC event along the same line of sight.

In our simulations we find that extinction in the foreground galaxies
does not severely affect the detectability of multiple lensed events
nor the ability to derive the flux-ratio between the images. Clearly,
multi-band observations of the SNe will be important in order to
correct for the different amounts of extinction of the images. At the same
time, the data could provide important results on the dust properties
of the foreground lensing galaxies, similar to the studies done with
multiple imaged quasars at $z_d\lsim 1$ \citep{falco}.  Unlike
quasars, CC events have a very deliberate color evolution along their
lightcurves. In general, from the moment of shock-breakout onwards,
the atmospheres of CC SNe expand and become cooler and
redder. This signature not only helps in the relative timing of these
events, but also allows us to measure the differential extinction due
to the varying amounts of dust along the lensed paths to the SN
quite well. With 3 or more filters one could make a measurement of
both the total extinction relative to the bluest event in addition to
the ratio of the total to selective extinction due to differences in
the dust properties. Furthermore, one could enhance this method by
taking a spectrum of the SN (any of the lensed events would do)
at a given epoch and through spectrum synthesis derive the true,
unextinguished spectral energy distribution of the event
[see e.g. \citep{mitchell,baron}].

\section*{Acknowledgements}
It is a pleasure to acknowledge E.~Linder for careful reading of the 
manuscript and important suggestions.
We thank G.~Bernstein, A.~Kim and the rest of the SNAP collaboration
for useful discussions. AG is a Royal Swedish Academy Research Fellow
supported by a grant from the Knut and Alice Wallenberg Foundation.
This work was supported by a NASA LTSA grant to PEN and by resources of
the National Energy Research Scientific Computing Center, which is
supported by the Office of Science of the U.S. Department of Energy under
Contract No. DE-AC03-76SF00098.   

\end{document}